\documentclass[conference]{IEEEtran}

\usepackage{amsthm,amssymb,graphicx,multirow,amsmath,color,amsfonts}
\usepackage[update,prepend]{epstopdf}
\usepackage[noadjust]{cite}
\usepackage[latin1]{inputenc}
\usepackage{bbm} 
\usepackage{pdfpages}
\usepackage{tabulary}
\usepackage{multirow}
\usepackage{comment}
\usepackage{array}
\usepackage{tabularx}

\usepackage{amsmath,amssymb}

\let\svbibcite\bibcite
\def\bibcite#1#2{\svbibcite{#1}{#2}}
\makeatletter
\let\svbiblabel\@biblabel
\def\@biblabel#1{\svbiblabel{#1}}
\makeatother

\begin{document}
\title{  Blocking Probability Analysis for 5G New Radio (NR) Physical Downlink Control Channel}    

\author{\IEEEauthorblockN{ Mohammad Mozaffari$^1$, Y.-P. Eric Wang$^1$, and Kittipong Kittichokechai$^2$}\vspace{-0.25cm}\\
	\IEEEauthorblockA{
		\small  $^1$ Ericsson Research, Santa Clara, CA, USA,  Emails:{\{mohammad.mozaffari, eric.yp.wang\}@ericsson.com}.\\
	\small  $^2$ Ericsson Research, Sweden,  Email: {kittipong.kittichokechai@ericsson.com}.\vspace{-0.1cm}}
		
	}\vspace{-0.72cm}
\maketitle\vspace{-0.2cm}
\vspace{-0.4cm}
\begin{abstract}
	The 5th generation (5G) new radio (NR) is designed to support a wide range of use cases, requirements, and services from enhanced mobile broadband (eMBB) to ultra-reliable low-latency communications (URLLC). NR signals are designed to meet stringent requirements, and in particular the physical downlink control channel (PDCCH) which requires careful consideration. In this paper, we provide a comprehensive analysis on the PDCCH blocking probability in a network where multiple users need to be scheduled for receiving downlink control information. We evaluate the performance of blocking probability in terms of various network parameters including number of users, size of the control resource set (CORESET), and PDCCH aggregation levels (ALs). Our analysis reveals fundamental tradeoffs and key insights for efficient network design in terms of PDCCH blocking probability.

\end{abstract} \vspace{0.3cm}

\section{Introduction}

The fifth generation (5G) new radio (NR) access technology,  introduced in Release 15 of the 3rd generation partnership project (3GPP), enables offering unique services for mobile broadband and ultra-reliable low-latency communications (URLLC) \cite{NR3GPP, NRthenew,5Gwireless}. With its deployment flexibility, wide range of spectrum availability, and ultra-lean design, 5G NR is able to effectively serve a variety of use cases with stringent requirements on data rate, latency and energy efficiency.  NR has been designed to operate at frequency range 1 (FR1)  from 410 MHz to 7.125 GHz and  frequency range 2 (FR2) from 24.25 GHz to 52.6 GHz. In addition, NR introduces unique features such as flexible numerology (e.g., subcarrier spacing and slot duration) and dynamic time division duplex (TDD), thus making it suitable for various deployment scenarios.  

Meanwhile, NR physical channels and signals are designed in a way to meet the 5G performance requirements. For instance, compared to long-term evolution (LTE), several enhancements have been made in designing synchronization signals and physical downlink control channel (PDCCH).  PDCCH carries downlink control information  (DCI) which plays a key role in downlink (DL) and uplink (UL) scheduling, as well as other aspects such as power control, slot format indication, and preemption indication. Ensuring a robust performance for PDCCH requires careful considerations. One key system performance evaluation metric is the PDCCH blocking probability which indicates the percentage of user equipments (UEs) that cannot be  scheduled by the network for receiving the DCI. Furthermore, the blocking probability impacts the latency which is a critical metric in many 5G use cases. Achieving a desired system performance requires minimizing the blocking probability. Note that blocking probability is a function of various network parameters such as number of UEs, size of the Control Resource Set (CORESET), PDCCH aggregation levels (ALs), and scheduling strategy. Therefore, in order to guarantee a minimum blocking probability, there is a need for in-depth evaluations of the impact of network parameters on the blocking probability.

\subsection{Related work on NR PDCCH}

In \cite{takeda2019}, the authors provide an overview of the 5G NR PDCCH by discussing physical layer structure of PDCCH, monitoring schemes, and DCI aspects. In \cite{Chen2}, the link-level performance of NR PDCCH is evaluated in terms of the block error rate (BLER). The work in \cite{Hamidi} studies the search space design for NR PDCCH while considering UE's PDCCH blind decoding (BD) and channel
estimation capabilities.  In \cite{Braun}, an overview of NR PDCCH as well as enhancement techniques for search space design (in particular PDCCH hash function) are presented. Moreover, the performance of the proposed techniques in \cite{Braun} are evaluated in terms of PDCCH blocking probability.  While previous studies provide some specific results for PDCCH blocking probability, the literature lacks a comprehensive analysis on this metric considering a wide range of relevant network parameters.

\subsection{Contributions}
In this paper, we provide an in-depth analysis on the NR PDCCH blocking probability in a network with multiple UEs that need to be scheduled for receiving the PDCCH. In particular, we evaluate the impact of various parameters including number of UEs, CORESET size, PDCCH ALs and their distribution, number of PDCCH candidates, UE's capability, and scheduling strategy on the blocking probability. Our analysis demonstrates  inherent tradeoffs and design insights for efficient network design in terms of PDCCH blocking probability. Specifically, one can minimize the blocking probability  by properly adjusting the network parameters based on the scenario.  


 The rest of this paper is organized as follows. In Section II, we provide the an overview of NR PDCCH. In Section III, we present the system model. Results and discussions are presented in Section IV and conclusions are drawn in Section V.
 
 \section{Overview of NR PDCCH}
\begin{figure}[!t]
  	\begin{center}
  		\includegraphics[width=9cm]{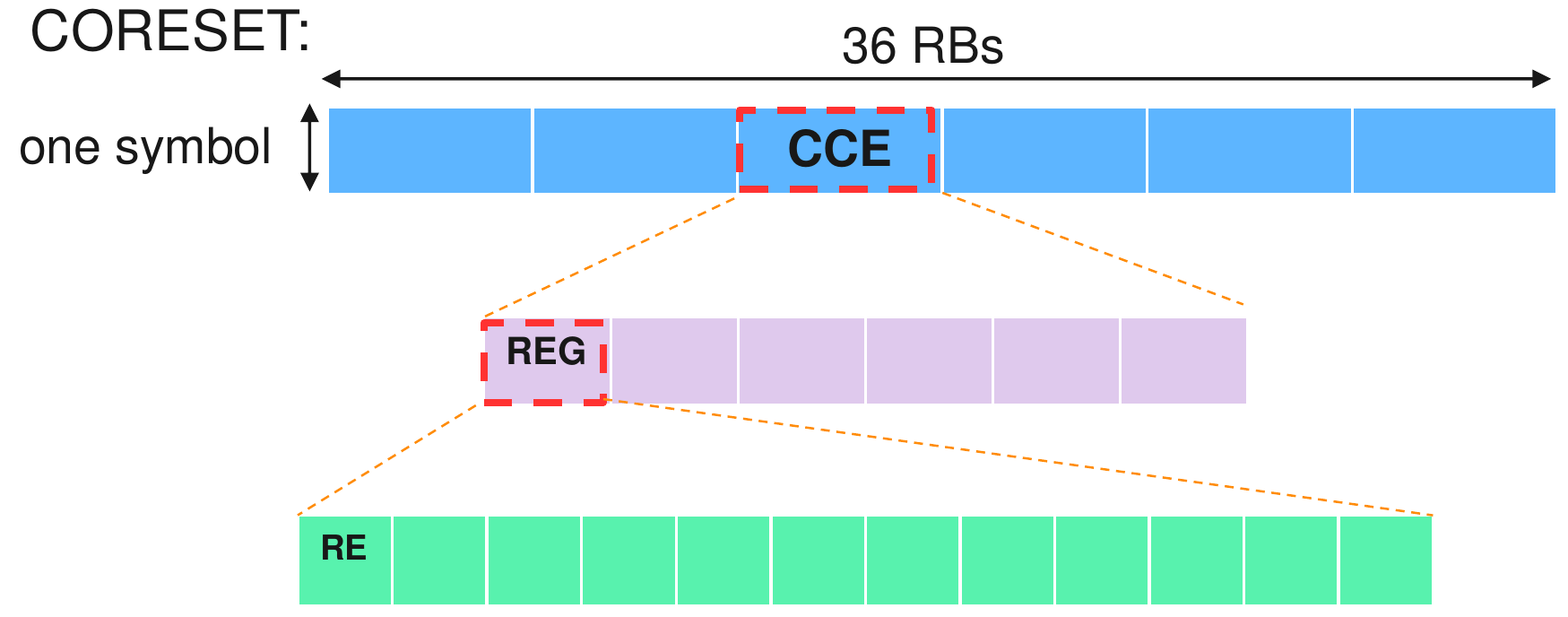}
  		\vspace{-0.2cm}
  		\caption{ An illustration of a CORESET with 36 RBs, one symbol (6 CCEs).}\vspace{-0.02cm}
  		\label{CORESET}
  	\end{center}
  \end{figure}

  PDCCH carries downlink control information for one or a group of UEs for several purposes such as DL scheduling assignment, UL scheduling grant, power control, and preemption indication. In NR, different DCI formats for different purposes are supported. Different DCI formats may or may not have different sizes. The size of a DCI format depends on the DCI fields that support specific features. 
  DCI is transmitted through PDCCH candidates which are located within  CORESETs. Each CORESET can  span over one, two, or three contiguous orthogonal frequency-division multiplexing (OFDM) symbols over multiple resource blocks (RBs), where each RB consists of 12 subcarriers. In the frequency domain, a CORESET spans over one or multiple chunks of 6 RBs \cite{ahmadi}.  A PDCCH candidate is carried by 1, 2, 4, 8 or 16 control channel elements (CCEs). Each CCE is composed of 6 resource element groups (REGs), and each REG is 12 resource elements (REs) in one OFDM symbol. Note that,  an RE is the basic resource unit in NR which consists of one subcarrier in one OFDM symbol.  In Figure \ref{CORESET}, we provide an illustrative example for  a CORESET with 36 RBs and one OFDM symbol consisting of 6 CCEs.

  Also, a REG bundle consists of multiple REGs where bundle size can be 2, 3, or 6, depending on the CORESET duration.  Each CORESET is associated with a CCE-to-REG mapping which can be interleaved or non-interleaved. In the non-interleaved case, all CCEs in an AL are mapped in consecutive REG bundles of the associated CORESET. In the interleaved case, REG bundles of CCEs are distributed on the frequency domain over the entire CORESET bandwidth.

\begin{figure}[!t]
	\begin{center}
		\includegraphics[width=9cm]{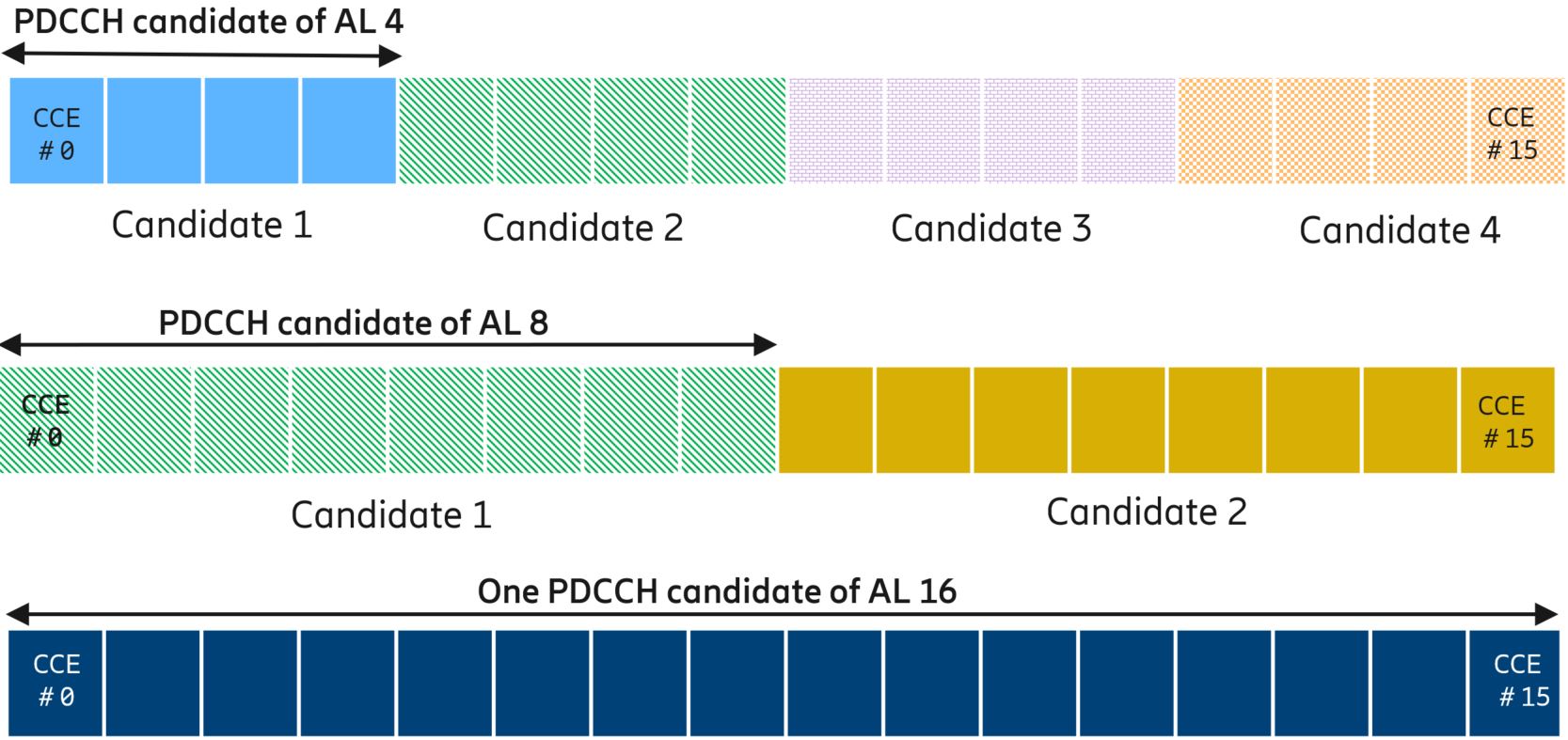}
		\vspace{-0.3cm}
		\caption{ An illustration of PDCCH candidates of ALs 4, 8, and 16 in a CORESET with 16 CCEs.}\vspace{-0.02cm}
		\label{Candidates}
	\end{center}
\end{figure}

In order to receive DCI, the UE needs to perform blind decoding as it is not aware of  the exact position of the PDCCH candidate used by the network. PDCCH candidates which need to be  monitored by UEs are configured using so-called search space (SS) sets with each SS being associated with one CORESET. In NR,  there are two types of SS: 1) common SS (CSS) set, commonly monitored
by a group of UEs, and 2) UE-specific SS (USS), monitored by a specific UE.  Within a search space configuration, various PDCCH monitoring parameters such as number of candidates, and possible number of CCEs in each candidate can be set \cite{TS_38.331}. The number of CCEs used for a PDCCH candidate is referred to as an aggregation level (AL). In NR, different aggregation levels can be used for PDCCH transmissions. Currently, possible NR PDCCH ALs are {1, 2, 4, 8, 16}. A higher AL provides better coverage and is more suitable for larger cells and extreme coverage scenarios, at the cost of more CCEs and consequently more time-frequency resources.   For each AL, the UE may need to monitor multiple candidates. 
In Figure \ref{Candidates}, we show an example of  PDCCH candidates with ALs 4, 8, and 16 in a CORESET composing of 16 CCEs.

To decode DCI, a UE performs blind decoding as it does not have explicit information about DCI size, AL, and the PDCCH candidate. In general, the number of blind decodes (BDs) depends on various factors such as the number of different DCI sizes, the number of ALs and the number of PDCCH candidates that need to be monitored for each AL. In order to limit the UE complexity and power consumption, there are limits on the maximum number of blind decoding and the number of non-overlapping CCEs for channel estimation per slot. The BD and CCE limits (for non-carrier aggregation) for 15/30/60/120 kHz subcarrier spacings (SCSs) are, respectively, 44/36/22/20 and 56/56/48/32 \cite{TS_38.213}. Next, we describe  our system model used for the blocking probability evaluations.

%

%

\section {System Model}	
Let $U$ be the number of UEs which need to be simultaneously scheduled by the network for receiving DCI. A gNB (i.e., 5G base station) uses a CORESET with $q$ RBs and $d$ symbol duration to schedule the UEs. In this case, the CORESET size in terms of number of CCEs is given by $C=\frac{q
\times d}{6}$. The CCEs within the CORESET are indexed from 0 to $C-1$. The gNB can use different set of ALs for scheduling the UEs. For each UE a suitable AL can be adopted based on several factors including the performance requirements and link quality. We use $p_L$ to denote the probability of using AL $L$ for the UEs in a cell. Specifically, $\mathbf{P}=[p_1, p_2, p_4, p_8, p_{16}]$ indicates the distribution  of ALs 1, 2, 4, 8, and 16.

 The position of different PDCCH candidates for each AL is determined using a hash function \cite{TS_38.213}. Let $l_{k,i}$ be the index of $(i+1)^{\text{th}}$ CCE of candidate $k$, where $ i \in \{0,..., L-1\}$. Therefore, CCE indices for candidate $k$ with AL $L$ (i.e., $L$ CCEs) are: $l_{k,0},..., l_{k,L-1}$. In a search space set  associated with a CORESET (with index $p$) in slot $t$, the CCE indices for PDCCH candidate $k$ are determined based on the following hash function (without carrier aggregation) \cite{TS_38.213}:
\begin{equation}\label{hash}
{l_{k,i}} = L\left[ {\left( {{Y_{p,t}} + \left\lfloor {\frac{{kC}}{{LM}}} \right\rfloor } \right)\it{mod} \left\lfloor {\frac{C}{L}} \right\rfloor } \right] + i,
\end{equation}	
where $\left\lfloor . \right\rfloor$ is the floor function and $\it{mod}$ represents the modulo operation. $M$ is the number of PDCCH candidates for AL $L$, and $ k \in \{0,..., M-1\}$ is the index of a PDCCH candidate with AL $L$.  Moreover, $Y_{p,t}$ is a constant value which is 0 for a CSS, and for a USS is given by \cite{TS_38.213}:
\begin{equation}\label{Yp}
{Y_{p,t}} = \left( {{A_p}{Y_{p,t-1}}} \right)\it{mod}\, (\text{65537}),
\end{equation}
where for the first slot (i.e., $t=0$), we have $Y_{p,-1}=n_{RNTI}=C_{RNTI} \ne 0$, with $C_{RNTI}$ being a unique identification number for each UE. $A_p=39827, 39829$, or $39839$, respectively, for $p$ $mod$ $3= 0, 1,$ \text{or}  $2$, where $p$ is the CORESET index.

From (\ref{hash}), we can see that the index of the first CCE of candidates with AL $L$ can be 0, $L$, $2L$, etc., as also illustrated in Figure \ref{Candidates} for $L=4$. 
	
The gNB can use different PDCCH candidates within the CORESET for scheduling different UEs. In this case, the blocking occurs for a UE when there is no fully free (i.e., non-overlapped) PDCCH candidates available for scheduling that UE. PDCCH blocking probability is defined as the probability that all PDCCH candidates configured for a UE to monitor are blocked  by candidates used by other UEs. That is,  the blocking probability is the ratio between the number of the blocked UEs over the number of all UEs that need to be scheduled, as written below:
\begin{equation}
B=\frac{\text{Number of blocked UEs}}{U},
\end{equation}
with $U$ being the total number of UEs to be scheduled. Note that the blocked UEs need to be scheduled at another PDCCH opportunity.   	
	
\begin{figure}[!t]
	\begin{center}
		\includegraphics[width=7cm]{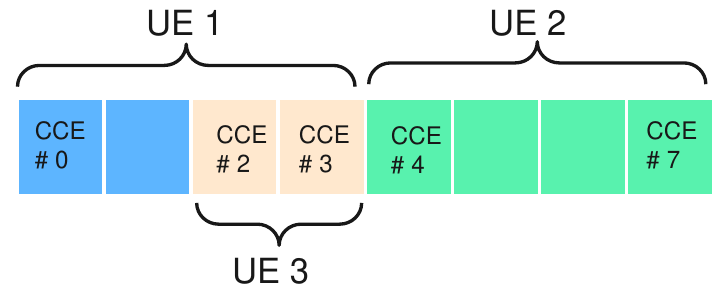}
		\vspace{-0.03cm}
		\caption{ Example of PDCCH blocking in a CORESET.}
		\label{Block}
	\end{center}
\end{figure}

As an example provided in Figure \ref{Block}, UE 2 (AL 4) is successfully scheduled while there is no non-overlapped candidates available for UE 1 (AL 4) and UE 3 (AL 2), thus one of them will be blocked. In this case, the blocking probability is $B=1/3$. In general, the PDCCH blocking probability is a complicated function of various parameters including number of UEs, CORESET size, ALs and their distribution, the number of candidates for each AL, and UE capability in terms of supported BD and CCE limits. Moreover, in a general case, there is no closed-form expression for the PDCCH blocking probability. Next, we investigate the impact of various parameters on the PDCCH blocking probability.

\section{Simulation Results and Analysis}
In this section, we provide simulation results for blocking probability evaluations while analyzing the effect of different parameters. Specifically, we investigate the impact of number of UEs, CORESET size, number of candidates, ALs and their distribution, UE capability, and scheduling strategy on the blocking probability. We focus on a USS and Monte Carlo simulations are performed over 10000 iterations.

\subsection{Impact of Number of UEs}
\begin{figure}[!t]
	\begin{center}
		\includegraphics[width=9cm]{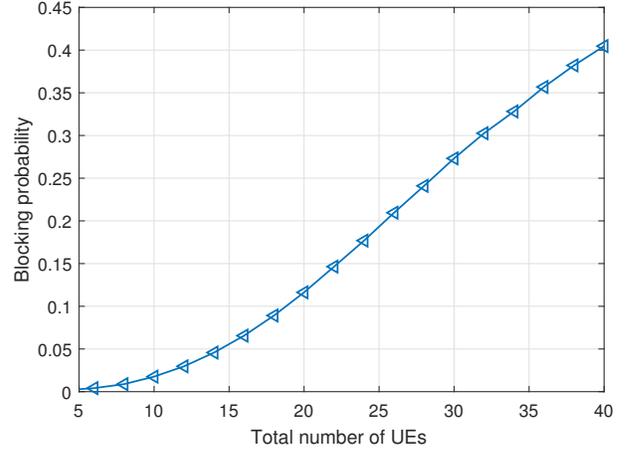}
		\vspace{-0.3cm}
		\caption{ Blocking probability versus number of  UEs to be scheduled.}\vspace{-0.02cm}
		\label{UE_number}
	\end{center}
\end{figure}

In order to evaluate the effect of number of UEs to be scheduled ($U$) on the blocking probability, we consider a CORESET of size 54 CCEs (e.g., a CORESET with 108 RBs and 3 symbols). Also, we consider ALs [1, 2, 4, 8, 16], with distribution [0.4, 0.3, 0.2, 0.05, 0.05]. For each UE, the number of PDCCH candidates for ALs [1, 2, 4, 8, 16] are, respectively, [6, 6, 4, 2, 1]. In Figure \ref{UE_number}, we show how the blocking probability varies by changing the number of UEs. As expected, the blocking probability increases when the number of UEs increases. Since more UEs are scheduled within a given CORESET, there will be a higher probability that the gNB does not find an available PDCCH candidate for a UE, thus resulting in a higher blocking probability. For example, Figure \ref{UE_number} shows that by doubling the number of UEs from  15 to 30, the blocking probability increase from 0.06 to 0.27, corresponding to an increase by a factor of 4.5.  

\subsection{Impact of CORESET Size}
\begin{figure}[!t]
	\begin{center}
		\includegraphics[width=9cm]{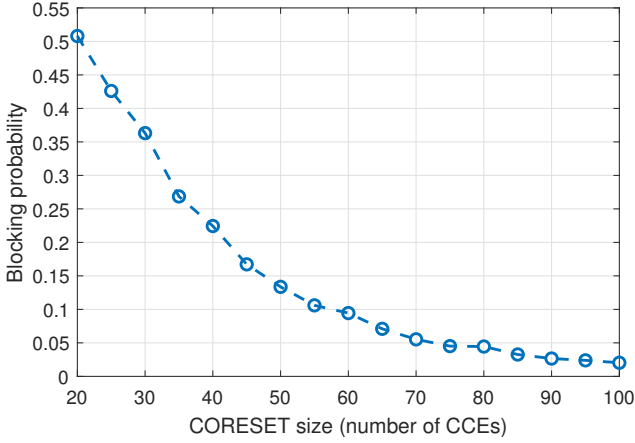}
		\vspace{-0.3cm}
		\caption{ Blocking probability versus CORESET size (number of CCEs).}\vspace{-0.02cm}
		\label{CORESET_num}
	\end{center}
\end{figure}

The CORESET size can significantly affect the blocking probability. Figure \ref{CORESET_num} shows  the blocking probability as a function of the CORESET size for $U=20$ UEs. As we can see, the blocking probability is decreasing by increasing the CORESET size. With a larger CORESET more CCEs and PDCCH candidates are available for scheduling the UEs. In addition, the scheduler has more flexibility for allocating PDCCH candidates to the UEs. From Figure \ref{CORESET_num} we can see that the blocking probability can reduced from 0.36 to 0.1 by increasing the number of CCEs in the CORESET from 30 to 60. Note that the impact of further increasing the CORESET size is minimal as almost all UEs can be successfully scheduled.

\subsection{Impact of Number of PDCCH Candidates}
\begin{figure}[!t]
	\begin{center}
		\includegraphics[width=9cm]{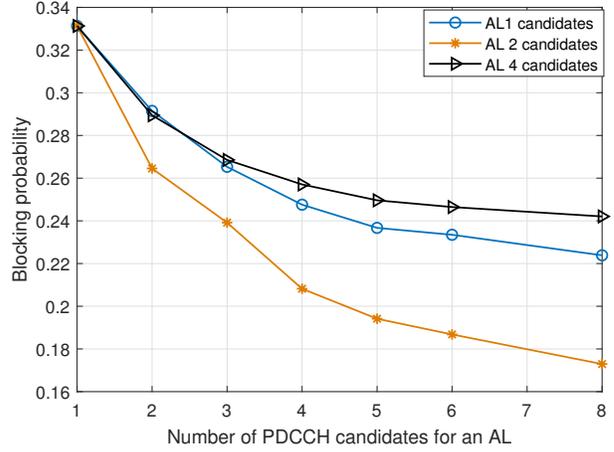}
		\vspace{-0.3cm}
		\caption{ Blocking probability versus the number of PDCCH candidates for an AL (20 UEs and CORESET size 54 CCEs).}\vspace{-0.02cm}
		\label{Candidate_num}
	\end{center}
\end{figure}

The number of PDCCH candidates for different ALs is another important factor. In NR, the number of PDCCH candidates can be configurable  for each aggregation level among $\{0, 1, 2, 3, 4, 5, 6, 8\}$ in the USS \cite{ahmadi, TS_38.213}. Note that for each UE, the locations of candidates are determined based on (\ref{hash}) and (\ref{Yp}), thus, different UEs have different CCEs mapped to a candidate. Here, we separately evaluate the impact of number of candidates for AL 1, AL 2, and AL 4. To this end, we only change the number of candidates for one of the ALs, while setting the number of candidates for other ALs to 1. The AL distribution is [0.4, 0.3, 0.2, 0.05, 0.05] for ALs [1, 2, 4, 8, 16].  Figure \ref{Candidate_num} shows that increasing the number of PDCCH candidates for each AL results in a lower blocking probability. With more PDCCH candidates, the gNB has more flexibility to avoid overlapping between candidates of different UEs, thus reducing the blocking probability. For instance, by increasing the number of candidates from 2 to 6 in this figure, we can observe the blocking probability reduction of 20\%, 30\%, and 17\%, respectively, for ALs 1, 2, and 4. Also, by increasing the number of candidates in Figure \ref{Candidate_num}, we see a higher blocking probability reduction for AL 2, compared to ALs 1 and 4. This is because, considering the AL distribution, the overall impact of AL 2 on the blocking probability is more than that of ALs 1 and 4. We note that, while having more PDCCH candidates is beneficial for blocking probability reduction, it increases the number of BDs and CCE monitoring which can increase  the UE complexity and power consumption. This shows a tradeoff between blocking probability and UE complexity/power consumption when increasing the number of PDCCH candidates.

\subsection{Impact of ALs}
\begin{figure}[!t]
	\begin{center}
		\includegraphics[width=8.2cm]{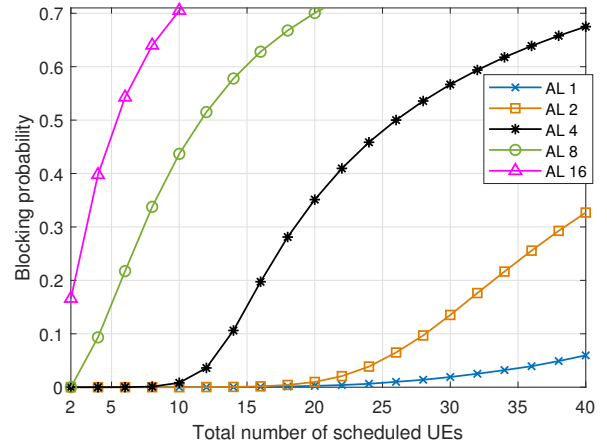}
		\vspace{-0.3cm}
		\caption{ Blocking probability for different ALs (CORESET size = 54 CCEs).}\vspace{-0.02cm}
		\label{ALs_impact}
	\end{center}
\end{figure}

As discussed earlier, a higher AL provides a better coverage at the cost of using more CCEs. Here, we primarily evaluate the effect of each AL on the blocking probability. Here, for the sake of evaluation, we consider using only one of the ALs among $\{1, 2, 4, 8, 16\}$ in each scenario. That is, in each scenario only one AL is used the probability of 1. Here, the number of PDCCH candidates for ALs 1, 2, 4, 8, and 16 are, respectively, 6, 6, 4, 2, and 1. For example, in case of AL 1, the network only configures 6 candidates for each UE to monitor (and other ALs are not monitored). As we can see, using a higher  AL leads to a higher blocking portability. Consequently, in order to guarantee a specific blocking probability, a smaller number of UEs can be simultaneously scheduled with a higher AL. The results in Figure \ref{ALs_impact} show that to maintain the blocking probability below 0.2, the maximum possible number of UEs to be scheduled with ALs 2, 4, 8, and 16 is 33, 16, 6, and 2, respectively.

\subsection{Impact of AL Distribution} \label{CoverageCond}
Note that the distribution of ALs can be determined based on the signal-to-interference-plus-noise ratio (SINR)  distribution of UEs (can be obtained e.g., from system-level simulations) and PDCCH link-level performance with different ALs. In fact, suitable ALs are used for UEs to meet the PDCCH performance requirements and one can find how ALs are distributed in a CORESET. For our evaluation in this section, we consider three scenarios corresponding to good, medium, and poor coverage. Specifically:
\begin{itemize}
\item Good coverage: most of UEs are in good coverage and require low ALs (i.e., ALs 1 and 2), with ALs distribution [0.5, 0.4, 0.07, 0.02, 0.01]. 

\item Medium coverage: most of UEs are in medium coverage and require medium ALs (i.e., AL 4),  with ALs distribution [0.05, 0.2, 0.5, 0.2, 0.05]. 

\item Extreme coverage: most of UEs are in poor coverage and require high ALs (i.e., ALs 8 and 16), with ALs distribution [0.01, 0.02, 0.07, 0.4, 0.5].
\end{itemize}
The CORESET size is 54 CCEs and the number of PDCCH candidates for ALs [1, 2, 4, 8, 16] are [6, 6, 4, 2, 1].

\begin{figure}[!t]
	\begin{center}
		\includegraphics[width=8.5cm]{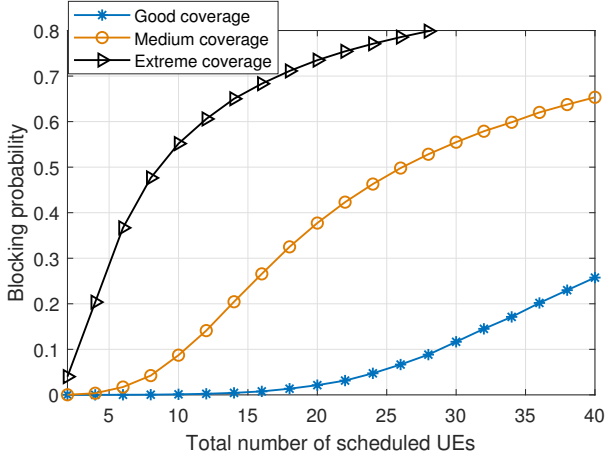}
		\vspace{-0.3cm}
		\caption{ Blocking probability for different AL distributions.}\vspace{-0.02cm}
		\label{AL_dist}
	\end{center}
\end{figure}

Figure \ref{AL_dist} shows that the blocking probability is lower for better coverage conditions. The ALs distribution depends on the coverage condition. As the coverage condition gets worse, it is more likely that higher ALs are used to meet the coverage requirements. This, in turn, increases the blocking probability. For example, for 20 UEs, the blocking probabilities for good, medium, and extreme coverage scenarios are 0.02, 0.38, and 0.72, respectively.  

\subsection{Impact of UE's Capability}
In this section, we analyze the impact of UE's capability in terms of BD/CCE limits on the blocking probability. In general, when BD/CCE limits is reduced, the UE can monitor a fewer number of PDCCH candidates per slot. This can also limit the scheduling flexibility and increases the blocking probability. For the evaluation of reduced BD limits, we consider the following cases, assuming that UE is configured with the maximum number of PDCCH candidates: 

\begin{itemize}
\item Reference case: we assume that the UE is configured to monitor [6, 6, 4, 2, 1] PDCCH candidates for ALs [1, 2, 4, 8, 16].

\item Reduced BD case $A$: the UE is configured to monitor  [3, 3, 2, 1, 1] PDCCH candidates for ALs [1, 2, 4, 8, 16]. In this case, the BD limit is reduced by around 50\% compared to the reference case.
 
\item Reduced BD, case $B$: the UE is configured to monitor  [1, 1, 1, 1, 1] PDCCH candidates for ALs [1, 2, 4, 8, 16]. In this case, the BD limit is reduced by around 75\% compared to the reference case.
\end{itemize}
We consider ALs distribution [0.4, 0.3, 0.2, 0.05, 0.05].

\begin{figure}[!t]
	\begin{center}
		\includegraphics[width=8.5cm]{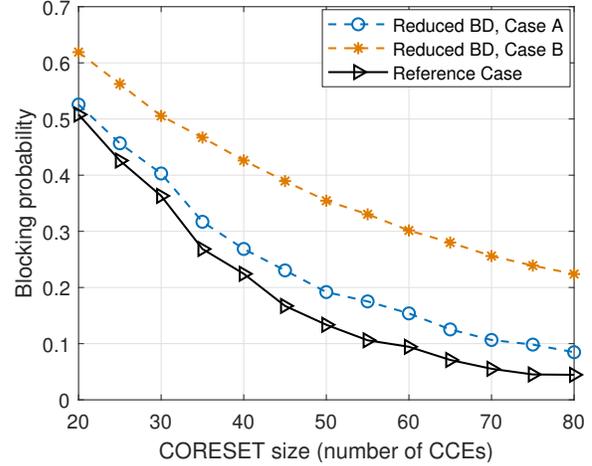}
		\vspace{-0.3cm}
		\caption{ Blocking probability for different blind decoding (BD) capabilities.}\vspace{-0.02cm}
		\label{BD_reduction}
	\end{center}
\end{figure}

Figure \ref{BD_reduction} shows that the blocking probability increases by reducing the BD limit. For instance, for a CORESET size of 54 CCEs, the blocking probability increase by factor of 1.9 and 3 when reducing the BD limit by 50\% and 75\% compared to the reference case.

\subsection{Impact of Scheduling Strategy}
Scheduling strategy is another impacting factor. In particular, it is important how the gNB allocates PDCCH candidates to different UEs. For instance, let us consider  two scheduling strategies:
\begin{itemize}

\item Strategy 1: scheduler allocates UEs from low-to-high ALs. That is, UEs with low ALs are scheduled first (this strategy is adopted in our evaluations). 

\item Strategy 2: scheduler allocates UEs from high-to-low ALs. That is, UEs with high ALs are scheduled first.
\end{itemize}

\begin{figure}[!t]
	\begin{center}
		\includegraphics[width=8.5cm]{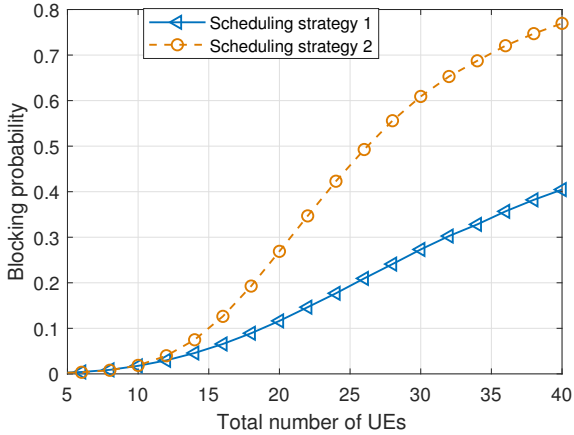}
		\vspace{-0.3cm}
		\caption{ Blocking probability for different scheduling strategies.}\vspace{-0.02cm}
		\label{Strategy}
	\end{center}
\end{figure}

Figure \ref{Strategy} shows that Strategy 1 outperforms Strategy 2 in terms of blocking probability. The reason is that Strategy 2 prioritizes UEs with high ALs that uses more CCEs, thus resulting  in a higher blocking probability compared to Strategy 1. As an example, in Strategy 2, a UE using AL 16 may block 16 UEs using AL 1. Note that the impact of scheduling strategy becomes more crucial as the number of UEs increases. According to Figure \ref{Strategy}, for a small number of UEs (e.g., 10) the two scheduling strategies have the same performance. However, when the number of UEs increases to 40, the blocking probability using Strategy 2 is 1.9 times larger than the case with Strategy 1, in the CORESET with 54 CCEs. It should be noted that the performance of different scheduling strategies is also dependent on the CORESET size.   

\subsection{Design Problem: Minimum CORESET Size for a Blocking Probability Target}
One key design problem is to determine the minimum CORESET size needed for meeting a blocking probability target. More specifically, given the number of UEs and the coverage condition, the network can properly determine the CORESET size to ensure the blocking probability does not exceed a specified threshold. 

\begin{figure}[!t]
	\begin{center}
		\includegraphics[width=8.5cm]{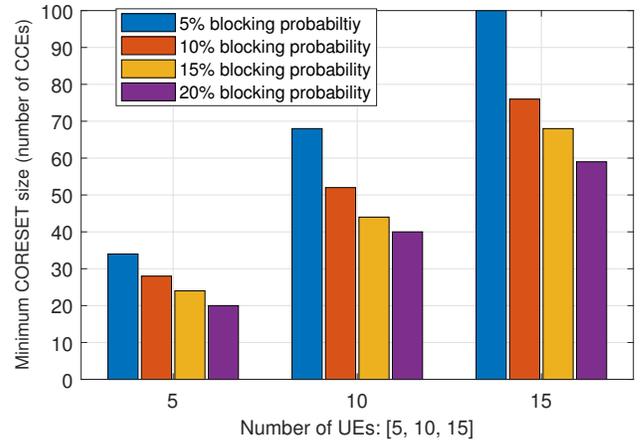}
		\vspace{-0.3cm}
		\caption{ Minimum required CORESET size for different number of UEs and blocking probability requirements.}\vspace{-0.02cm}
		\label{min_CORESET}
	\end{center}
\end{figure}

We consider the medium coverage condition presented in Section \ref {CoverageCond} and find the minimum CORESET size that ensures the blocking probability below certain thresholds. Figure \ref{min_CORESET} shows the minimum required CORESET size for {5, 10, 15} UEs and different blocking probability targets {5\%, 10\%, 15\%, 20\%}. Clearly, the CORESET size must increase when more UEs are scheduled and a smaller blocking portability target needs to be met. For example, comparing two cases: i) 5 UEs and 20\% blocking probability, and ii) 15 UEs and 5\% blocking probability requirement, shows that CORESET size for the later case needs to be 5 times larger than that of the former case (i.e., from 20 CCEs to 100 CCEs). While a larger CORESET is beneficial for UE scheduling, it may not be desired from spectral and energy efficiency perspective. Therefore, the network should properly select the CORESET size based on the requirements and deployment scenarios.

\section{Conclusions}	
In this paper, we have conducted a comprehensive  analysis on the NR PDCCH blocking probability in a network with multiple UEs that need to be scheduled for receiving the PDCCH. We have evaluated the impact of a wide range of parameters and design factors on the blocking probability. In particular, we have analyzed the effect of  number of UEs, CORESET size, PDCCH ALs and their distribution, PDCCH candidates, UE's capability, and scheduling strategy on the blocking probability. Our analysis along with simulation results have shown  fundamental tradeoffs and design insights for efficient network design in terms of PDCCH blocking probability.	In particular, based on the scenario, constraints, and system parameters (e.g., number of UEs, and CORESET size), one can adopt effective techniques to reduce the blocking probability. For instance, in a scenario with limited CORESET size and good coverage condition, efficient scheduling strategies and increasing the number of PDCCH candidates for small ALs can be effective for blocking probability reduction. \vspace{0.1cm}

\def\baselinestretch{1.04}
\bibliographystyle{IEEEtran}
\bibliography{referenceConf}

\end{document}